\documentclass[a4paper]{article}
\usepackage[abbr,agsm-eprint]{harvard}
\usepackage{hyperref}
\usepackage[reqno,intlimits]{amsmath}
\usepackage{amsthm}
\usepackage{amssymb}
\usepackage{graphicx}
\usepackage[american]{babel}
\begin{document}
\title{NEUTRINOS AND GRAVITATIONAL WAVES FROM COSMOLOGICAL GAMMA-RAY
BURSTS}
  \author{Giulio Auriemma\\ Universit\`a degli Studi della Basilicata, Potenza, Italy}
  \date{Invited paper presented at the \emph{``Vulcano
Workshop 2002''}, Vulcano, Italy, 20-25 May, 2002.} \maketitle
\renewcommand{\min}{\mathrm{min}}
\renewcommand{\max}{\mathrm{max}}
\newcommand{\apjl}{Astrophys. J. Lett.}
\newcommand{\apj}{Astrophys. J.}
\newcommand{\aj}{Astron. J.}
\newcommand{\apss}{Astroph. and Space Sci.}
\newcommand{\mnras}{Monthly. Not. R. Astron. Soc.}
\begin{abstract}
Cosmological gamma ray bursts are very likely powerful sources of
high energy neutrinos and gravitational waves. The aim of this
paper is to review and update the current predictions about the
intensity of emission in this two forms to be expected from GRB's.
In particular a revised calculation of the neutrino emission by
photohadronic interaction at the internal shock is obtained by
numerical integration, including both the resonant and the
hadronization channels. The detectability of gravitational waves
from individual bursts could be difficult for presently planned
detectors if the GRB's are beamed, but it is possible, as we have
proposed in a paper two years ago, that the incoherent
superimposition of small amplitude pulse trains of GW's impinging
on the detector, could be detected as an excess of noise in the
full VIRGO detector, integrating over a time of the order of one
year.
\end{abstract}
\newpage
\section{Introduction}
The Gamma Ray Bursts (GRB) are intense flashes of soft
$\gamma$-rays, serendipitously discovered by the Vela test ban
treaty \cite{Klebesadel73} and immediately confirmed by the Soviet
Konus satellites \cite{Mazets74}. The phenomenology of GRB's has
been discussed in several reviews (see e.g.
\citeasnoun{Djorgovski:2001du} and references therein). The most
important breakthrough in the GRB's observations has been the
discovery in 1997 of a long lasting X-ray ``afterglow'' of the
burst GRB970228 by the Beppo-Sax Italian-Dutch satellite (Costa et
al. 1997), which has allowed the identification of its host galaxy
\cite{vanParadijs:1997wr,Sahu:1997vp}, making the study of GRB's a
``multi-wavelengths adventure'' \cite{2000RvMA...13..129K}.

According to a widely accepted (and acceptable) interpretation,
 the prompt $\gamma$-rays burst and the delayed afterglow, are both powered
 by the dissipation of the kinetic energy of a relativistically expanding
 ``fireball'' (or jet), whose primal cause is not known.
This mechanism, qualitatively suggested by
\citeasnoun{1978MNRAS.183..359C}, has been refined by
\citeasnoun{1986ApJ...308L..43P} and put in its final form of a
standard model by \citeasnoun{1994ApJ...430L..93R} (for a recent
review see e.g. \citeasnoun{2002ARA&A..40..137M} and references
therein). The obvious expectation is that such a powerful highly
relativistic explosion should be accompanied by emission of
neutrinos and gravitational waves, thus becoming the herald of the
new astronomy.  \par The aim of this paper is to review and update
the current predictions about the intensity of emission in this
two forms to be expected from GRB's. In the following I will first
review in \S\ref{sect:energy} and \S\ref{sect:large} the essential
quantitative information on the energetics and the large scale
distribution of the sources, which has been obtained from prompt
and afterglows observations. In \S\ref{sect:neutrino} I present a
new calculation of the photohadronic neutrino emissivity in the
fireball/jet system from which predictions about the observability
of the diffuse neutrino background can be derived. In
\S\ref{sect:GW} I reexamine the possibility that a similar
stochastic background could be observed as an additional noise
component in the long term integration of the signal from
gravitational wave detectors.

\section{Energetics}\label{sect:energy}

The total energetics involved in the GRB's is the most important
parameter for my subsequent emissivity estimate.  A direct
estimate of the rest frame energy output can be obtained only for
bursts with known redshift. According to a compilation by J.
Greiner, available at
\texttt{http://www.mpe.mpg.de/$\sim$jcg/grbrsh.html}, the number
of GRB with reliable redshift measurement of the host galaxy
redshift, is about 20.
\par The intrinsic isotropic bolometric (from 0.1 keV to 10 MeV)
energy release, estimated from the fluence of 17 burst in the 30
keV -2 MeV range, spans from $5.6\times10^{51}\;\mathrm{erg}$
 (GRB990712) to $2\times10^{54}\;\mathrm{erg}$ (GRB990123), with a
 median of
$\langle\mathcal{E}_{\gamma}^{iso}\rangle=1.8\times10^{53}\;\mathrm{erg}$
and a dispersion of $0.8\;\mathrm{dex}$(r.m.s.)
\cite{2001AJ....121.2879B}.  However, it is to be emphasized that
this value applies only to GRBs with measured redshift. Therefore
observational biases in burst detection and redshift
determination, might obscure the true underlying average energy
output of the typical GRB. However, keeping in  mind this caveat,
one can assume that the kinetic energy input to the expanding
relativistic fireball is of the order of
$$\langle\mathcal{E}_{0}^{iso}\rangle\approx
1\,\left(\frac{20\%}{\eta_\gamma}\right)\,M_\odot\,c^2$$ where
$\eta_\gamma$ is the fraction of the fireball kinetic energy that
is converted into $\gamma$rays \cite{2001ApJ...557..399G}.

The afterglow light curve of many bursts shows an achromatic
break, currently interpreted  as the evidence for a beamed
emission \cite{Sari:1999mr,Rhoads:1999wn}. In this case the
emitted $\gamma$-ray energy $\mathcal{E}_\gamma^{jet}$  is
obviously reduced by the factor \cite{Frail:2001qp}
$\mathcal{E}_\gamma^{jet}=f_{jet}\,\mathcal{E}_\gamma^{iso}$ where
$f_{jet}=1-\cos\theta_{jet}$. In a recent paper
\cite{Bloom:2003eq} present a complete sample of 29 gamma-ray
bursts (GRBs) for which it has been possible to determine temporal
breaks (or limits) from their afterglow light curves.
Incorporating realistic estimates of the ambient density and
propagating error estimates on the measured quantities, in
agreement with the previous analysis of a smaller sample, the
derived jet opening angles of those 29 bursts result in a narrow
clustering of geometrically corrected gamma-ray energies about
$\langle\mathcal{E}^{jet}_{\gamma}\rangle = 1.33 \times 10^{51}
\;\mathrm{ergs}$, with a burst-to-burst variance of
$0.34\;\mathrm{dex}$. \par Three consequences can be drawn, if
this is true:\begin{enumerate}
    \item The central engines of GRBs release energies of the order of
$$\mathcal{E}_{0}\approx \mathrm{few}\times 10^{-2}\,\left(\frac{20\%}
{\eta_\gamma}\right)\;M_\odot\,c^2$$ This value is indeed
comparable to the kinetic energy output of ordinary SNe,
suggesting a connection between the two phenomena, also supported
by the observation of bumps in the afterglows light curves
\cite{Bloom:2003se};
    \item The large spread in observed fluence and peak luminosity of GRBs
    is mostly due to a spread in the opening angle of the beam;
    \item Only a small fraction of GRB's are visible to a
given observer and the true GRB's rate is
$$R_{GRB}^{jet}=R_{GRB}^{iso}\times\langle f_{jet}^{-1}\rangle$$ In
practice from Table 2 of the paper by \citeasnoun{Bloom:2003eq}
one can guess that the true GRB's rate should be about 300 times
larger than the observed one.
\end{enumerate}

It is clear that such a radical descoping of the energetics
associated with the GRB's explosion, has a fundamental impact on
the physical interpretation, but has little or no impact, at the
first order, in the predictions of their high energy neutrino
emission. In fact we expect that the aperture of the beam of
neutrinos emitted by photohadronic processes will be the same of
the beam of $\gamma$-rays emitted by electrons.

On the contrary the gravitational wave emission by GRB, if
present, will be in any case isotropic. We observe that the
beaming could reduce the probability of detecting individual
events, because the amount of energy to be radiated is strongly
reduced. But this should have no impact on the prediction of a
possible cosmological background originated by the GRB, because in
case of beaming we should assume that the rate of events is
increased by the same factor. Therefore we have that the total
energy released is in the case of beaming will be in the average
$$
    \mathcal{E}_{GW}^{jet}\times
R_{GRB}^{jet}=\mathcal{E}_{GW}^{iso}\times R_{GRB}^{iso}
$$

As we said already the above given estimate can be applied, only
to the bursts with observed afterglow. Since afterglows have been
observed only for long lasting bursts with duration $T_{90\%}\ge
2\;\mathrm{s}$, it could be argued that the short bursts could be
powered by a totally different central engine, whose energetic
could be on a lower scale. Some indications in this direction is
given by the fact that the observed fluence of the short bursts is
significantly lower then the one of long bursts
\cite{Tavani:1998ps}. However one could argue back that the
difference in the average fluence could be due to a markedly
different large scale distribution of the two classes of bursts.
In the absence of a any direct distance estimator, only
statistical indicators of the spatial distribution of GRB can be
used to test this hypothesis. A recent analysis of the 4B burst
catalog \cite{Schmidt:2001uz} shows that the short bursts have
essentially the same characteristic peak luminosity of the long
ones, but their local space density is around three times lower.
This applies to bursts shorter then 0.25 s, therefore the total
energy output of the short bursts should be 1-2 order of magnitude
lower then the one of the long burst.

\section{Large scale distribution}\label{sect:large}

\begin{figure}[h!]
  \centering
\includegraphics[width=0.7\textwidth]{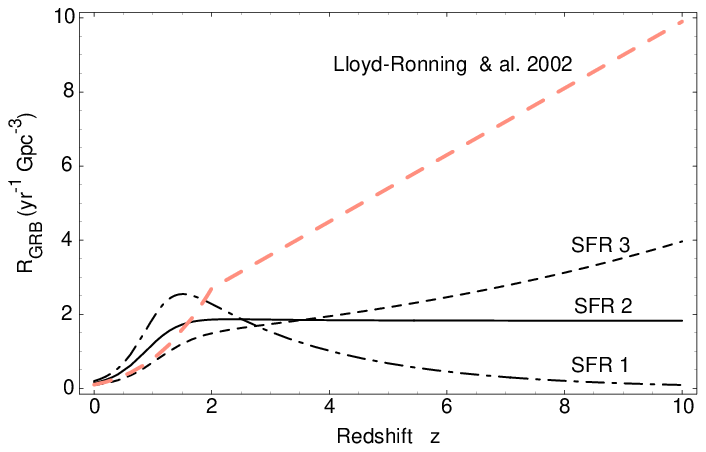}
  \caption{}\label{fig:rate}
\end{figure}

The common wisdom about the large scale distribution of the bursts
is that, being a phenomenon originated by the collapse of short
lived massive stars, they should track closely the Star Formation
Rate (SFR) cosmological distribution. In a recent paper
\citeasnoun{Porciani:2000} have used three different SFR derived
from other astronomical data to fit the observed log N-Log P
distribution for the bursts. Their results imply about 1-2 GRB's
per million SN type II and a characteristic isotropic luminosity
of the bursts in the range
$3-20\times10^{51}\;\mathrm{ergs}\,\mathrm{s}^-1$. The best fitted
models of the rate per unit comoving volume are shown in Fig.\
\ref{fig:rate}.

The luminosity-variability relation\cite{Fenimore:2000vs} is a new
entirely empirical connection, which has been proposed in analogy
to the well understood luminosity-period relationship of the
cepheids. Having calibrated over a sample of 8 GRB's with known
redshift, the authors have applied this fit to derive the
redshifts for 220 bursts.  Applying non-parametric statistical
techniques to the 220 Gamma-Ray Burst (GRB) redshifts and
luminosities derived in this way, it has been proposed
\cite{Lloyd-Ronning:2001bg} that there exists a significant
correlation between the GRB intrinsic luminosity and redshift,
which can be parameterized as $(\Omega_\gamma/4\pi)\,L_\gamma
\propto (1+z)^{1.4\pm 0.5}$, where z is the burst redshift.  In
addition, this analysis supports a co-moving rate density of GRBs
that continues to increase for $(1+z)>10$.
We have plotted in the Fig.\ \ref{fig:rate} the different possible
large scale distribution of the GRB's, which is compatible with
the observed log N-log P distribution.

\section{High energy neutrinos}\label{sect:neutrino}

In the general framework of the relativistic fireball model of
GRB's, the prompt $\gamma$-ray emission is originated by
synchrotron emission  of relativistic electrons accelerated at the
internal shocks, while the afterglow is due to the interaction of
the external shock with the ambient material \cite{Piran:1999bk}.
The radius of the inner shock (or shocks in case of multi-peaked
emission) should be of the order of $R_{s}\approx 2\,
\Gamma_b^2\,c\,\Delta t$, where $\Delta t$ is the smallest
timescale observed in the burst emission, usually in the range
1-10 ms.\par A lower limit to the bulk Lorentz factor of the
expanding fireball can be derived from the $\gamma\gamma\to
e^+\,e^-$ self absorption in the burst at the highest observed
$\gamma$-ray energy $\epsilon_\gamma^{max}\simeq
100\;\mathrm{MeV}$. In fact given the threshold of the reaction
$2\,m_e\,c^2\simeq 1\; \mathrm{MeV}$ we must have $\Gamma_b\ge
\epsilon_\gamma^{max}/\sqrt{2\,m_e\,c^2}\simeq 100$. A more
detailed calculation \cite{2001ApJ...555..540L} shows that the
optical depth is $\tau_{\gamma\gamma}\propto
\Gamma_b^{-4+2\alpha}$, where $\alpha$ is the low energy spectral
index of the prompt $\gamma$-ray emissioin. Therefore a bulk
Lorentz factor $\Gamma_b\gtrsim 300$ is required for the emission
of $\sim100\;\mathrm{MeV}$ $\gamma$-rays.

It has been shown \cite{1995ApJ...453..883V,1995PhRvL..75..386W}
that the high magnetic fields and relativistic velocities of the
internal shocks,  can accelerate protons up to
$10^{19}-10^{20}\;\mathrm{eV}$, corresponding to the range of the
UHE cosmic rays. From first principles one can derive the rate of
acceleration from the relativistic shock
$\dot\epsilon_p/\epsilon_p\simeq \frac{e}{m_p}\, B\,\Gamma_b^2$,
thus even in a very short time, smaller than the coherence time of
the $\gamma$-ray emission, a huge maximum energy
$$\ln\left(\frac{\epsilon_p^{max}}{\epsilon_p^{min}}\right)\simeq
\frac{e}{m_p} B\,\Gamma_b\,\Delta t$$ can be reached.
\par However the shock acceleration mechanism is efficient only when
the radius of the shock is larger then the Larmor radius
$r_L\simeq c\, \epsilon_p/e\, B$ of the accelerated particle.
Therefore the more stringent limit will be $r_L\lesssim R_s$, that
assuming $R_s\simeq 2 \,c\, \Gamma_b^2\,\Delta t$ gives
$\epsilon_p^{max}\lesssim e\,B\, \Gamma_b^2\,\Delta t$. But even
in this case one can get $\epsilon_p^{max}\approx
10^{19}\;\mathrm{eV}$ with field of the order $\sim
1\;\mathrm{T}$. The typical fields required in the burst for
electron synchrotron emission peaked in $\sim 100\;\mathrm{keV}$
range are $\gg 1\;\mathrm{T}$ \cite{2000ApJ...543..722L}.
Therefore the maximum energy imparted to the proton will be
limited, in practice, by proton synchrotron radiation and
inelastic scattering over the radiation field. The latter
interaction will copiously produce pions, that subsequently decay
into neutrinos and TeV $\gamma$-rays.
\begin{figure}[h!]
  \centering
\includegraphics[width=\textwidth]{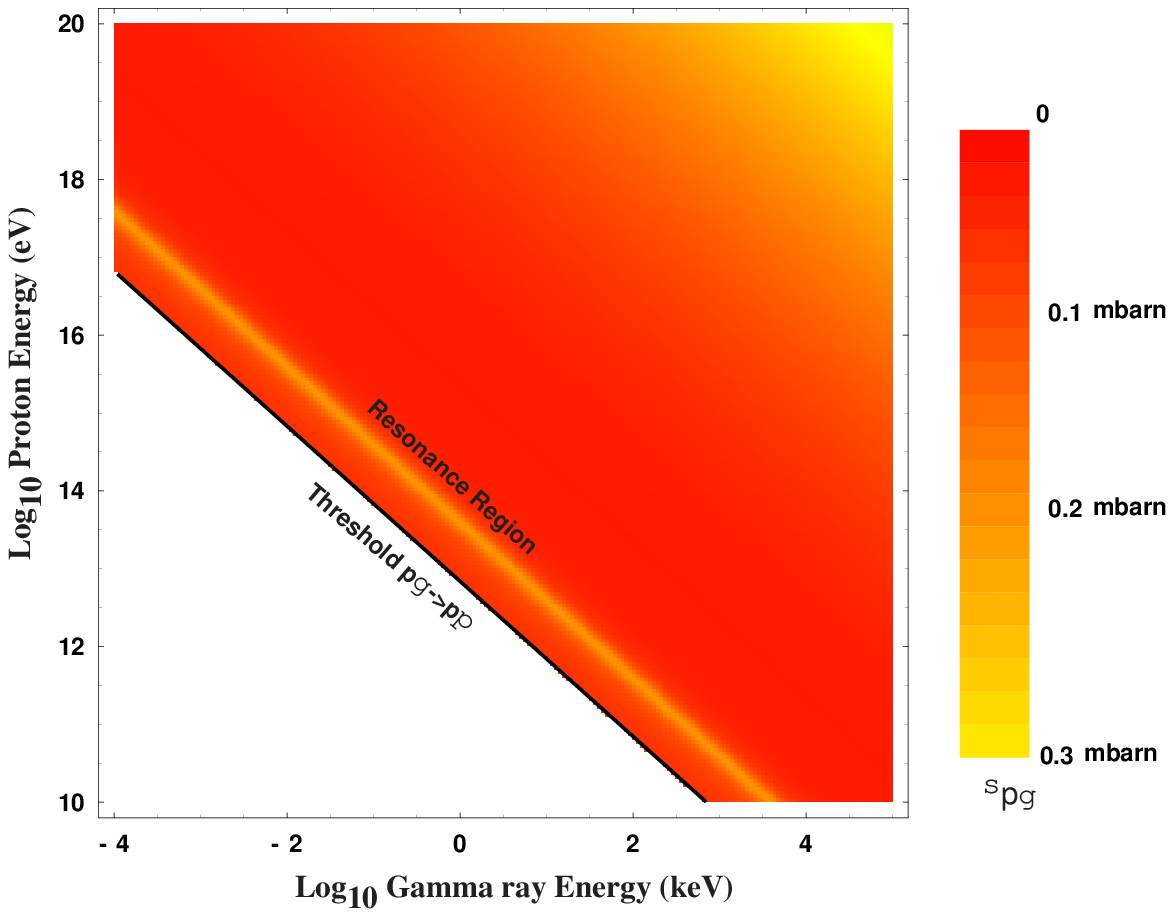}
  \caption{}\label{fig:kine}
\end{figure}
The pion yield for the reaction $p\,\gamma\to \pi^\pm\,X$ at the
internal shock has been calculated, under simplifying assumptions,
by several authors (see e.g. \citeasnoun{Guetta:2003wi} and
references therein). All this previous calculations have included
only the photohadronic production near threshold, where the photon
scatters quasi-elastically  with the nucleon that is pushed into
an excited state corresponding to $N^+$ and $\Delta^+$ baryonic
resonances:
\begin{equation}\label{eq:neutrino1}
    p\, \gamma\to \left\{\begin{array}{c}
  N^+ \\
  \Delta^+ \\
\end{array}\right\}\to \left\{\,
\begin{array}{ll}
    n\,\pi^+\\
    p\,\pi^0\\
\end{array}
\right.
\end{equation} These resonances are produced
with high cross sections ($\sim 100-500\;\mu\mathrm{barn}$)
\cite{Hagiwara:2002fs}. In addition to these resonant channels,
pions will be also produced at intermediate energies by the
hadronization of the photon $\gamma\to q\bar q$.  The reaction
proceed in practice as
\begin{equation}\label{eq:neutrino12}
    \gamma p\to (q\bar q) p\to \mathcal{N}, \pi^\pm,\pi^0,\cdots
\end{equation}
The importance of considering both processes, and not only the
resonant channels, in photohadronic pion production calculations
has been already stressed in the literature \cite{Mucke:1998mk}.
In addition to the arguments given in that paper, one should
consider that the photons which interact via the resonant channel
should have energies in the range
\begin{equation}\label{eq:neutrino11}
   \Gamma_b^2
    \frac{M_{N^+,\Delta^+}^2-m_p^2}{4\epsilon_p}\le\epsilon_\gamma\le\epsilon_\gamma^{max}
\end{equation}
In Fig.\ \ref{fig:kine} we have plotted the cross section versus
the energies, in the observer's frame, of the photon and the
$\gamma$-ray. If the low-energy spectral index is $\alpha<0$,
protons with $\epsilon_p\ge 10^{13}\;\mathrm{eV}$ will produce
baryonic resonances scattering on photons with $\epsilon_\gamma\le
1\;\mathrm{keV}$. Therefore we expect that the resonant
photoproduction at high energies will depend strongly on the shape
of the $\gamma$-ray spectrum, in a range well below the observed
one.\par The theoretical expectation from the electron synchrotron
emission \cite{Sari:1999kj} is that the spectrum will level off
due to electron cooling at energies $\epsilon_\gamma\lesssim 100\;
\mathrm{eV}$ and be strongly suppressed by self-absorption for
$\epsilon_\gamma\lesssim10^{-3}\;\mathrm{eV}$. Following this
indication we have assumed  a photon energy distribution given by
the form empirically proposed by \citeasnoun{1993ApJ...413..281B},
modified to take into account the cooling:
\begin{equation}\label{eq:gamma2}
    \phi(\epsilon_\gamma)=A\,\left\{
\begin{array}{ll}
   E_c^\alpha\,e^{(\beta-\alpha)E_c/E_b} & \mathrm{if}\; \epsilon_\gamma\le E_c
   \\[6pt]
   \epsilon_\gamma^\alpha\,e^{(\beta-\alpha)\epsilon_\gamma/E_b} & \mathrm{if}\; \epsilon_\gamma<E_b \\[6pt]
    \epsilon_\gamma^\beta\,E_b^{\alpha-\beta}\,e^{\beta-\alpha} & \mathrm{if}\; \epsilon_\gamma\ge E_b \\
\end{array}
\right.
\end{equation}
in the range $10^{-3}\;\mathrm{eV}\le\epsilon_\gamma\le
100\;\mathrm{MeV}$ and zero outside this range. The constant $A$
has been chosen in order to normalize to 1 the function
$\psi(\epsilon_\gamma)$ in the above given range. Under these
assumptions the photon density at the source will be
\begin{equation}\label{eq:gamma3}
    n_\gamma\simeq \frac{L_\gamma\,\Delta t}{4\pi R_s^2 \Delta R\,
\langle\epsilon_\gamma\rangle}\, \phi(\epsilon_\gamma)
\end{equation}
where $R_s\simeq 2 c\,\Gamma_b^2\,\Delta t$.

The full expression for the m.f.p. of a proton with energy
$\epsilon_p$ in a radiation field with photon energies
$\epsilon_\gamma^{min}\le \epsilon_\gamma\le\epsilon_\gamma^{max}$
in observer's frame is given by
\begin{equation}\label{eq:neutrino2}
    \lambda_{p\gamma}^{-1}=\frac{\Gamma_b}{4\,\epsilon_p}
    \int_{m_p^2}^{mp^2+4\epsilon_p\epsilon_\gamma^{max}/\Gamma_b^2}
\sigma_{p\gamma}(s)\int_{\max\left[\epsilon_\gamma^{min},\Gamma_b\frac{s-m_p^2}{4
\epsilon_p}\right]}^{\epsilon_\gamma^{max}}
\frac{dn_\gamma}{d\epsilon_\gamma}
\,\frac{d\epsilon_\gamma}{\epsilon_\gamma}\,\,ds
\end{equation}
where, $\sigma_{p\gamma}$ is given by the sum of Breit-Wigner
resonances and hadronization cross sections
\begin{equation}\label{eq:neutrino3}
    \sigma_{p\gamma}(s)\simeq\sum_{k=N^+,\Delta^+}\,
    \sigma^{max}_k\,\frac{\frac{1}{4}\,\Gamma_k^2}{(\sqrt{s}
    -M_k)^2+\frac{1}{4}\Gamma_k^2}+\sigma_h (s)
\end{equation}
, we can calculate the average number of inelastic interactions
per proton
\begin{equation}\label{eq:neutrino4}
\eta_{\pi}(\epsilon_p)=\frac{\Delta
R}{\lambda_{p\gamma}(\epsilon_p)}=\frac{L_\gamma\,
\lambda_{p\gamma}^{-1}}{16 \pi c^2\,\Gamma_b^4\,\Delta
t\,\langle\epsilon_\gamma\rangle}
\end{equation}
The probability of having one or more proton-photon inelastic
scattering is $P_{\pi}=1-\exp(\eta_\pi)$ which is $P_\pi\approx
\eta_\pi$ only when $\eta_\pi\ll 1$. This is not the case,
specially in the hadronization channel, because the inelasticity
of the interaction is of the order of $k_{inel}\simeq40\%$.
Therefore one has to take into account that the proton actually
starts an small hadronic cascade in the considered layer.
\begin{figure}[h!]
  \centering
\includegraphics[width=\textwidth]{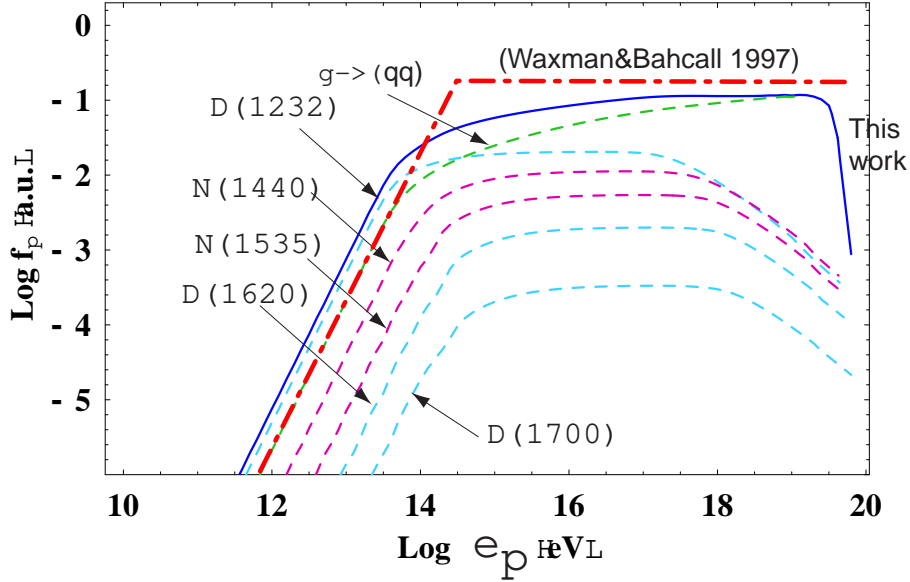}
  \caption{Fraction of energy lost by protons to pions, assuming
  a photon energy distribution given by Eq.\ \eqref{eq:gamma2},with low-energy spectral
  index $\alpha=-1$, high-energy index $\beta=-2.25$,
  a break energy $E_{b}=511\;\mathrm{keV}$, and a cooling energy $E_c=100\;\mathrm{eV}$.
  For the fireball's parameters it is assumed $L_\gamma=10^{52}\;\mathrm{erg}\,\mathrm{s}^{-1},
  \Gamma_b=300\;\mathrm{and}\;\Delta t=10\;\mathrm{ms}$.}\label{fig:fpion}
\end{figure}
We have integrated numerically Eq.\ \eqref{eq:neutrino2} including
all the resonant channels and the continuum hadronization cross
section, approximated by a polynomial logarithmic fit
\cite{Cudell:1999tx}. We have also approximately taken into
account the possibility of multiple interactions, simply summing
over the histories. In practice we have considered that the pions
of a given energy $\epsilon_\pi$ could have been produced either
by a proton of energy $\epsilon_p$ or by a proton which suffered
$n$ previous scattering, with $n\le
\log(\epsilon_p/\epsilon_\pi)/\log(1-k_{inel})$. Thus we have in
practice:
\begin{equation}\label{eq:neut6}
    \eta'_\pi(\epsilon_\pi)=\sum_{n=1}^{n_{max}}\,\xi^{-\gamma\,n}
    \,\bar n_\pi(\xi^{n}\epsilon_\pi)\,
    \left\{1-\exp\left[\eta_\pi(\xi^{n}\epsilon_\pi)\right]\right\}
\end{equation}
where is $\xi=1/(1-k_{inel})$, $\gamma$ the spectral index of the
proton spectrum, and $\bar n_\pi$ is either the average
multiplicity of the multiple pion production for the hadronization
channel or the branching ratios in the case of the resonant
channel. In Fig.\ \ref{fig:fpion} we report the fraction of energy
lost by protons to pions
$f_\pi=\eta'_\pi\times\langle\epsilon_\pi/\epsilon_p\rangle$ for
comparison with other calculations. As can be seen the result is
in reasonable agreement with the first estimate given by
\citeasnoun{Waxman:1997ti} which is shown as an heavy dot-dashed
line in the same figure.
\begin{figure}[h!]
  \centering
\includegraphics[width=\textwidth]{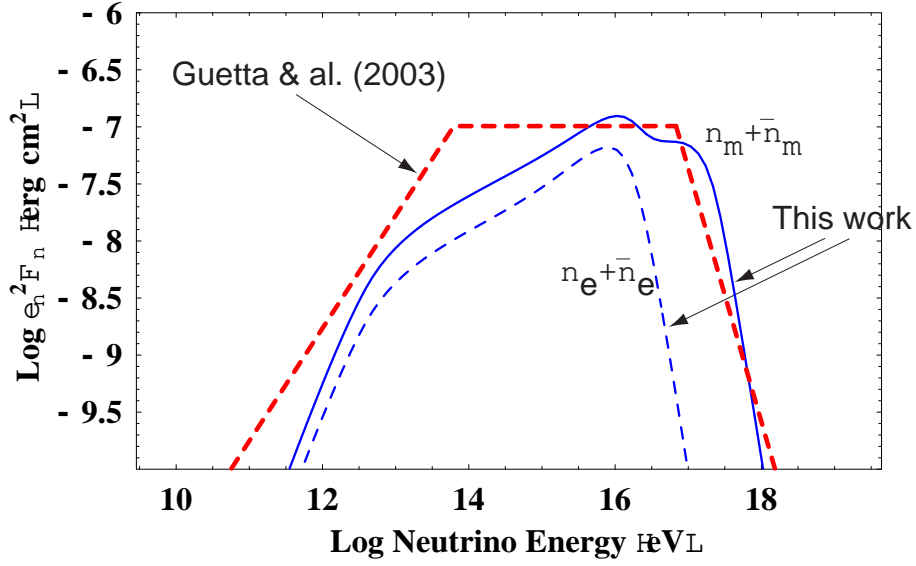}
  \caption{Typical neutrino spectrum for a burst with
  $\mathcal{E}_p=2.2\times10^{52}\;\mathrm{erg}$ with spectrum
  $\propto\epsilon_p^{-2}$ for
  $10^{10}\;\mathrm{eV}<\epsilon_p<10^{20}\;\mathrm{eV}$,
  exploding at $z=1$ in a Universe with
$H_0=71\;\mathrm{km}\,\mathrm{s}^{-1}\,\mathrm{Mpc}^{-1}$,
  $\Omega_M=0.3$ and $\Omega_\Lambda=0.7$ and $w=-1$ (see text).
  The fireball parameters are the same of Fig.\ \ref{fig:fpion}.
This hypothetical burst (assuming $\mathcal{E}_e=\mathcal{E}_p$
and $f_\gamma=0.04$) would have a fluence at Earth in the 20-2000
keV band of $3\times10^{-7}\;\mathrm{erg}\,\mathrm{cm}^{-2}$.
}\label{fig:typ_nu_spect}
\end{figure}
In Fig.\ \ref{fig:typ_nu_spect} we report the energy distribution
for the neutrino produced in the decay chain
\begin{equation}\label{neut5}
\pi\to\nu_\mu\,\mu\quad\mathrm{and}\quad\mu\to \bar \nu_{\mu}\,e\,
\nu_e
\end{equation}
In both cases it is important to take into account the pion (and
muon) energy loss, before decaying. This has been done solving
analytically the coupled equations for decay and energy loss,
which gives as a result when the dominant energy loss is the
synchrotron emission:
\begin{equation}\label{eq:neut10}
    \eta_{\nu_{\mu}}(\epsilon_{\nu_{\mu}})=\frac{m_\pi}{2\,\epsilon_{\nu_{\mu}}^*}\,
    \int_{\frac{m_\pi}{2\,\epsilon_{\nu_{\mu}}^*}\,
    \epsilon_{\nu_{\mu}}}^{\epsilon_\pi^{max}}\,\eta'_\pi(\tilde\epsilon_\pi)\,
    \left(\frac{\epsilon_{\pi c}}{\tilde\epsilon_\pi}\right)^2
    e^{-\frac{\epsilon_\pi^2-\tilde\epsilon_\pi^2}{2\,\epsilon_\pi^2}
    \left(\frac{\epsilon_{\pi c}}
    {\tilde\epsilon_\pi}\right)^2
}\,\frac{d\tilde\epsilon_\pi}{\tilde\epsilon_\pi}
\end{equation}
where $\epsilon_{\nu_{\mu}}^*=(m_\pi^2-m_\mu^2)/2m_\pi$ is the
energy of the neutrino in the muon rest frame, and $\epsilon_{\pi
c}$ is the energy of the pion for which the rate of energy loss is
equal to the decay time. For $\epsilon_\pi\gg\epsilon_{\pi c}$
Eq.\ \eqref{eq:neut10} gives the expected behavior
$\propto(\epsilon_\pi/\epsilon_{\pi c})^{-2}$, while for
$\epsilon_\pi\approx\epsilon_{\pi c}$ a pile up peak is generated.
\begin{figure}[h!]
  \centering
\includegraphics[width=\textwidth]{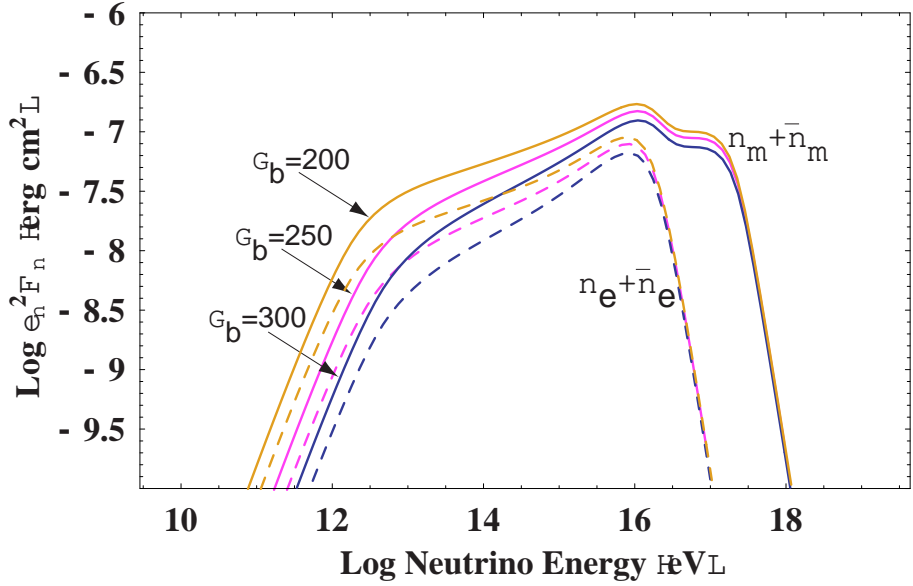}
  \caption{Neutrino spectrum for different bulk Lorentz factor
  keeping all the other parameters constant.}\label{fig:typ_nu_spect_comp}
\end{figure}
We can derive the spectrum of the muons from the same formula with
the only difference that the integration limits are in this case
$\frac{m_\pi\,\epsilon_\mu}{2\,\epsilon_\mu^*(1+\beta^*)}\le
\epsilon_\pi\le\frac{m_\pi\,\epsilon_\mu}{2\,\epsilon_\mu^*(1-\beta^*)}$
where $\epsilon_\mu^*=\sqrt{m_\mu^2+{\epsilon_{\nu_{\mu}}^*}^2}$
is the energy of the muon in the pion rest frame and $\beta^*$ its
velocity. In the muon decay  the energy loss before decay will be
more severe because the decay constant of the muon is longer
(implying $\epsilon_{\mu c}<<\epsilon_{\pi c}$), but we can use a
formula equivalent to Eq.\ \eqref{eq:neut10}. For the sake of
simplicity we have calculated the combined neutrinos and
anti-neutrinos flux, because in the multiple pion production it
can be assumed that positive and negative pions are equally
produced. The marked difference between the spectrum of muon
neutrinos and the one of the electronic neutrinos is due to the
fact that the first type of neutrinos are directly produced in the
pion's decay, while the electronic ones are produced only in the
muon's decay, which have suffered a larger energy loss. In Fig.\
\ref{fig:typ_nu_spect} we show for comparison the neutrino flux
calculated by \citeasnoun{Guetta:2003wi}, which is also in good
agreement with this result. In Fig.\ \ref{fig:typ_nu_spect_comp},
we report our prediction for the expected neutrino flux from the
photohadronic channel, for different values of the bulk Lorentz
factor. It is to be remarked that I do not find in the final flux
estimate the  strong dependance $\propto\Gamma_b^{-4}$ which is
argued by many authors (see e.g. \citeasnoun{Guetta:2003wi}). The
reason is that even if the average number of inelastic collision
$\eta_\pi$ calculated by Eq.\ \eqref{eq:neutrino4} is $\propto
\Gamma_b^{-4}$, the pion yield is $\eta'_\pi\propto\eta_\pi$ only
when $\eta_\pi\ll 1$. As we said above when it becomes
$\eta_\pi\gtrsim 1$ one should take into account the multiple
regenerative interactions, which take place at decreasing
energies, as has been taken into account using Eq.\
\eqref{eq:neut10}.
\begin{figure}[h!]
  \centering
\includegraphics[width=\textwidth]{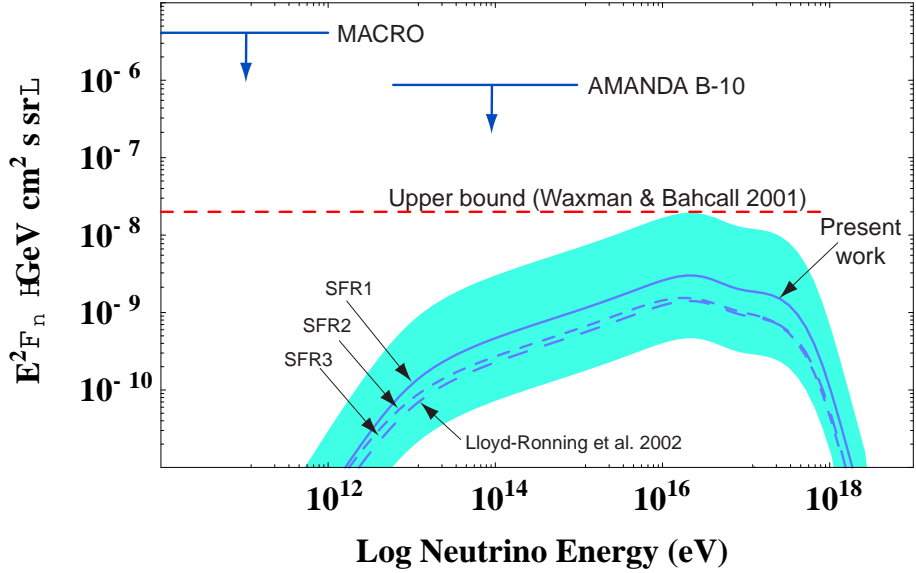}
  \caption{Diffuse neutrino flux for the different models of GRB's large scale distributions
  shown in Fig.\ \ref{fig:rate}, for a cosmological model with the same parameters of
  Fig.\ \ref{fig:typ_nu_spect}. The normalization of the curves
  has been obtained assuming an average total energy for accelerated protons in
  the burst rest frame of $\mathcal{E}_p=10^{52}\;\mathrm{erg}$.
  The band shown in this figure corresponds to
  $\pm0.8\;\mathrm{dex}$ variation of this average.
  }\label{fig:neut_back}
\end{figure}
In Fig.\ \ref{fig:neut_back} we report the diffuse cosmological
background flux of neutrinos obtained from our emissivity estimate
and the different large scale distribution models discussed in
\S\ref{sect:large}. The neutrino fluence at Earth of a burst with
redshift $z$ has been estimated as
\begin{equation}\label{neut20}
    \epsilon_\nu^{2}\,\mathcal{F}_\nu(z,\epsilon_\nu)=\frac{\eta'_\nu((1+z)\epsilon_\nu)\,
  }{4\pi\,(1+z)d_L^2(z)}
    \,\frac{\mathcal{E}_p}{\ln(\epsilon_p^{max}/
    \epsilon_p^{min})}
\end{equation}
where $d_L(z)$ is the cosmological luminosity distance,
$\mathcal{E}_{p}$ the total accelerated proton energy with
spectral index $-2$, and $\eta'_\nu$ is given by Eq.\
\eqref{eq:neut10}. The diffuse flux of neutrinos has been
calculated integrating
\begin{equation}\label{eq:gamma44}
    \Phi_\nu(\epsilon_\nu)=\int_0^{\infty}\,\frac{R_{GRB}(z)}{(1+z)}\,
    \mathcal{F}_\nu(z,\epsilon_\nu)\,\frac{dV}{dz}\,dz
\end{equation}
where
\begin{equation}\label{eq:gamma4}
   \frac{dV}{dz}=\frac{c\,d_L^2(z)}{(1+z)^2\,H(z)}
\end{equation}
being for Friedmann-Robertson-Walker metric \cite{Hagiwara:2002fs}
\begin{equation}\label{eq:gamma5}
    H(z)=H_0\,\sqrt{\Omega_M (1+z)^3+\Omega_K (1+z)^2+\Omega_V (1+z)^{-3(1+w)}}
\end{equation}
with $\Omega_K=1-\Omega_M -\Omega_V$.
\begin{figure}[h!]
  \centering
\includegraphics[width=\textwidth]{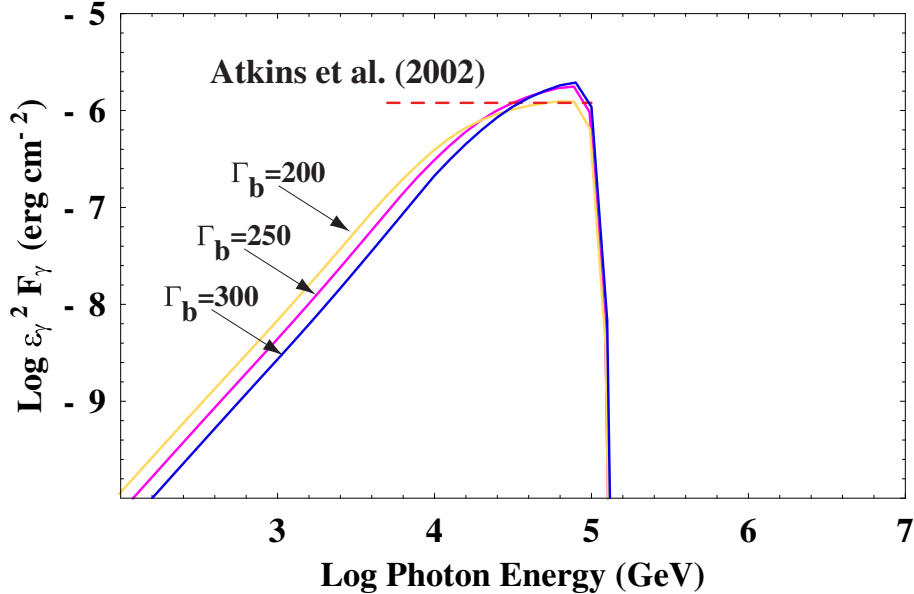}
  \caption{The predicted VHE $\gamma$-ray fluence of a very nearby bursts
  ($z=0.01$).}\label{fig:VHEgamma}
\end{figure}

Finally in Fig.\ \ref{fig:VHEgamma} we report our prediction for
the fluence of the $\gamma$-rays with $\epsilon_\gamma
>100\;\mathrm{GeV}$ produced by neutral pion decays. This mechanism has been
discussed in the literature (for a recent review see
\citeasnoun{Fragile:2002mz} and references therein) as a possible
source for VHE $\gamma$-rays potentially detectable by ground
based experiments. It is well known that only one statistically
significant detection has been reported until now by
\citeasnoun{Atkins:2002ws} using the Milagrito detector. This type
of signal is expected to be detectable only for very energetic and
nearby GRB's for two reason: 1) the strong self absorption due to
$\gamma\gamma\to e^+e^-$ interactions with the soft $\gamma$-ray
radiation field in the source, and 2) the absorption of the VHE
photons by the cosmic microwave background photon due to the same
process. The first process is responsible for the suppression of
the emission below $\approx 1-10\;\mathrm{TeV}$, while the second
originates the cut-off at 1 PeV.

\section{Gravitational waves}\label{sect:GW}
Prompt gravitational waves emission has been predicted from GRB
since 1993 by \citeasnoun{1993ApJ...417L..17K} at a level to be
detected by upcoming gravitational wave-experiments. As we have
discussed in \S\ref{sect:energy}, the estimate of the kinetic
energy required to power a beamed $\gamma$-ray burst is of the
order of $\mathcal{E}_0\gtrsim 0.01\,M_\odot\,c^2$, much larger
then the total nuclear binding energy of few-$M_\odot$'s stars.
Therefore in any case the energetic of the bursts is compatible
only with the collapse of a several solar masses object into a
black hole \cite{1993AAS...182.5505W}. Any reasonable assumption
regarding angular momentum then leads to a massive accretion disk
that flows, on a viscous time scale, into a black hole that very
rapidly approaches the Kerr limit. The dissipation the rotational
and gravitational energy within the geometry of a thick torus is
very likely to lead to jets, possibly by the purely
electromagnetic mechanism proposed by
\citeasnoun{1977MNRAS.179..433B}. The energy irradiated in this
case as gravitational waves \cite{1985PhRvL..55..891S} is
\begin{equation}\label{eq:GW1}
    \mathcal{E}_{GW}=\epsilon^4\,\left(\frac{m}{M_{BH}}\right)^2\,M_{BH}\,c^2
\end{equation}
 where $m$ is the mass of the torus, $M_{BH}$ is the mass of the black hole, and $\epsilon$
 is the ellipticity of the torus. It is worth noticing that if the
 jets are accelerated by Poynting flux \cite{Drenkhahn:2001ue} the amount
of kinetic energy going into the burst $\mathcal{E}_0$ and the
energy radiated in gravitational waves are strictly related,
because the GW luminosity will be $L_{GW}\propto\langle
\stackrel{\cdots}{Q_{ij}}\, \stackrel{\cdots}{Q^{ij}}\rangle$,
being $Q$ the quadrupole moment of the mass distribution of the
torus, while the Poynting flux will be $L_{e.m.}\propto \langle
\stackrel{\cdots}{M_{ij}}\, \stackrel{\cdots}{M^{ij}}\rangle$
being in this case $M$ the quadrupole moment of the magnetic field
embedded in the torus. Therefore we assume that the energy
radiated as gravitational radiation is
$\mathcal{E}_{GW}=\eta_{GW}\,\mathcal{E}_{0}$ where
$\eta_{GW}\lesssim 1$ is an unknown parameter.

The energy flux (\emph{viz.} energy per unit surface and unit
time) of GW produced by a burst of intrinsic luminosity (in
gravitational waves) $L_{GW}$ at a red shift $z$ is by definition
$\Phi_{GW}=L_{GW}/4\pi\,d_{L}^2(z)$ if $d_{L}$ is the luminosity
distance. In order to convert this flux into an adimensional
amplitude we can use the classical formula
\cite{Shapiro-Teukolsky83}:
\begin{equation}
    \Phi_{GW}=\frac{c^{3}}{16\pi G}\,\left\langle \dot
    h^{2}_{+}+\dot h^{2}_{\times}\right\rangle
    \label{eq:single2}
\end{equation}
where the average is taken over several wavelengths. The amplitude
of the signal produced depends from the direction and the beam
pattern of the detector. In the best case we have integrating over
time and applying the Parseval' s theorem
\begin{equation}
    \int_{-\infty}^{+\infty} \omega^2\,\tilde h^{2}(\omega)\,d\omega=
    \frac{16\pi G}{c^{3}}\frac{(1+z)\,E_{GW}}{4\pi\,d_{L}^2(z)}
    \label{eq:single3}
\end{equation}
where $\tilde h^{2}(\omega)$ will be the Rayleigh power of the
signal as a function of the frequency $\omega$. In order to
estimate the order of magnitude of the amplitude of the GW signal
we do not need a detailed shape of the spectral power density of
the signal, but only the knowledge of the firsts two moments of
the distribution,\emph{ viz.} $\bar\omega$ and $\Delta\omega$. In
fact we can recast Eq.\ (\ref{eq:single3}) in the form
$$\left(\bar
\omega^{2}+\Delta\omega^{2}\right)\,\int_{-\infty}^{+\infty}
\tilde h^{2}(\omega)\,d\omega= \frac{16\pi
G}{c^{3}}\frac{(1+z)\,E_{GW}}{4\pi\,d_{L}^2(z)}$$ It is remarkable
that rather natural physical assumption on the first and second
moment of the unknown Rayleigh power distribution $\tilde
h^{2}(\omega)$ can be made. In fact we can assume that the first
moment will be twice the keplerian angular velocity of the
marginally stable orbit around the BH, namely
\begin{equation}\label{eq:GW2}
    \bar \omega=2\sqrt{\frac{2 G_N\,M_{BH}}{\beta^3 r_S^3}}\simeq
    2\,\pi\,10^3\,\left(\frac{7\,M_\odot}{M_{BH}}\right)\,\left(\frac{2.5}{\beta}\right)^{3/2}
    \;\mathrm{rad}\,\mathrm{s}^{-1}
\end{equation}  where $r_{S}$ will be the Schwarzschild radius of
the collapsed object, which is function only of the black hole
mass, and $\frac{1}{2}\le\beta\le 3$ depends on the angular
momentum of the BH, being minimal for maximally  rotating BH, and
maximal for non-rotating ones.

The second moment will be the r.m.s. bandwidth of a wave packet
that can be estimated $\Delta\omega\approx 2\pi/\Delta t$ where
$\Delta t$ is the timescale of the emission in the comoving frame.
As we have seen in \S\ref{sect:neutrino} the $\gamma$-ray emission
has a variability over a timescale $\Delta t\simeq
10\;\mathrm{ms}$, therefore it is rather natural to assume that
the timescale of the GW emission by the torus will be of the same
order of magnitude. In this case we have, if $c\,\Delta t\gg
r_{S}$, for the peak amplitude
\begin{equation}\label{eq:single4}
    d_L(z)\,\tilde h^{peak}  \approx \sqrt{\frac{2\,\beta^3\,G_N^3}{\pi\,c^9}}\,M_{BH}\,
    \sqrt{E_{GW}\,\Delta t}
\end{equation}
The expected amplitude for typical values of beamed bursts is
$$d_L\,\tilde h^{peak}\approx
  10^{-26}\,
  \left(\frac{M_{BH}}{7\;M_\odot}\right)\,
  \left(\frac{\mathcal{E}_{0}}{6.5\times10^{51}\;\mathrm{erg}}\,\frac{\eta_{GW}}{0.01}\,
  \frac{\Delta t}{10\;\mathrm{ms}}\right)^{\frac{1}{2}}
  \;\mathrm{Gpc}/\sqrt{\mathrm{Hz}}$$
 compatible with our previous estimate \cite{Auriemma:2001rp}. The predicted signal
 by \citeasnoun{2002ApJ...580.1024C} for a ``gentle chirp'' spectral model, is
 well in agreement with the above value, when scaled by a factor $10^5$
 to take into account the different assumption on the distance and the total energy,
 which are in that paper respectively $d_L=100$ Mpc and $\mathcal{E}_{GW}=3.75\times10^{53}$ erg.
The current upper limit for the search for emission of GW in
correlations with GRB with bar detectors
\cite{Tricarico:2001gf,Astone:2002jz} has given the upper limit
$h_{RMS}\le 1.5\times 10^{-18}$. It is clear from this estimate
that the probability of detecting a single burst, even under
optimal  is very low, unless $\eta_{GW}\gg1$. However even if the
individual burst could not be detected, it is to be remarked that
in case of beaming the rate of explosion could be very large
($\approx 1500$ per day), therefore the stochastic accumulation of
signal integrated over a long time could emerge from the noise, as
we have shown in our previous paper \cite{Auriemma:2001rp}. The
same type of calculation, but with more optimistic assumptions on
the energetic of the source, as we said before, has been repeated
by \citeasnoun{2002ApJ...580.1024C}.
The energy flux of GW produced at Earth from a cosmological
distribution of sources can be calculated integrating over the
cosmological distribution, as we have done for the neutrino in the
Eq.\ \eqref{neut20}, namely
\begin{equation}\label{eq:single1}
    \Phi_{GW}^{diff}=\int_0^{z_{max}}\,\frac{R^{jet}_{GRB}(z)}{(1+z)}\,
    \frac{(1+z)\,E_{GW}}{4\pi\,d_{L}^2(z)}\,\frac{dV}{dz}\,dz
\end{equation}
where $dV/dz$ is given by Eq.\ \eqref{eq:gamma4}. Applying again
Eq.\ (\ref{eq:single2}) we have:
\begin{equation}
    \frac{c^{3}}{16\pi G}\,\left\langle \dot
    h^{2}_{+}+\dot h^{2}_{\times}\right\rangle=\frac{c}{4\pi}
    \int_{0}^{z_{max}}\,\frac{R^{jet}_{GRB}(z)\,E_{GW}}{(1+z)^2\,
    H(z)}\,dz
    \label{eq:back2}
\end{equation}
On the L.H.S. of this equation we have the amplitude of the wave
that invests at a certain instant the detector. We have seen in
the previous section that each of this burst will not have a
detectable intensity, but if we average over an observation time
$T$ long compared to the GW burst duration but short compared to
Hubble time scale (typically one year) we have a signal
\begin{equation}
    \frac{1}{T}\int_{-\infty}^{+\infty}\,\omega^{2}\, \tilde
    h^{2}(\omega)\,d\omega=\frac{8\,G}{c^2}
    \int_{0}^{z_{max}}\,\frac{R^{jet}_{GRB}(z)\,E_{GW}}{(1+z)^2\,
    H(z)}\,dz
    \label{eq:back3}
\end{equation}
that will be detectable if the power spectral density is greater
then the power spectral of the noise, averaged over the same
observation time.
\begin{figure}[h!]
  \centering
\includegraphics[width=\textwidth]{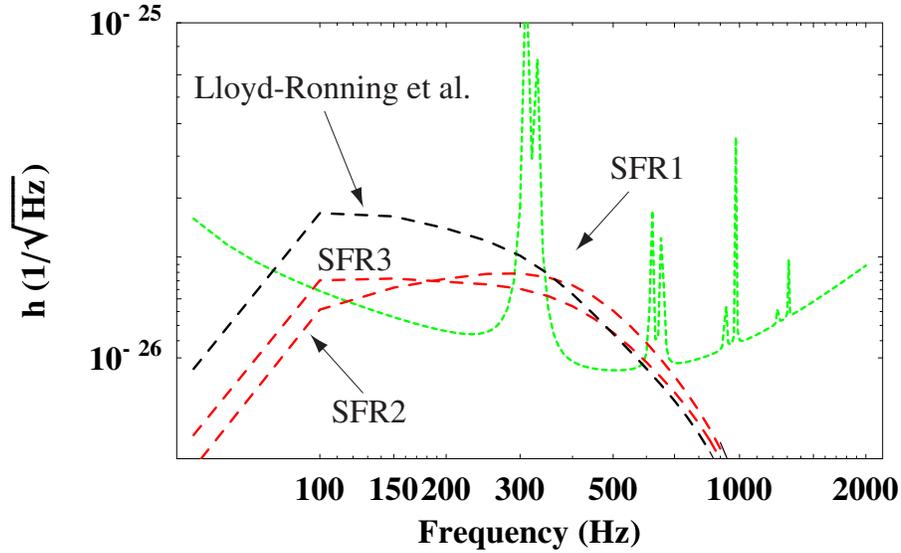}
  \caption{Stochastic signal integrated over one year from
cosmological GRB ($\eta_{GW}=0.01$) with large scale distributions
of Fig.\ \ref{fig:rate}, multiplied by an average beaming factor
$\langle f_{jet}^{-1}\rangle=300$. The dotted line is the noise
estimated in the VIRGO experiment (see text).}\label{fig:fig2_GW}
\end{figure}
The uncorrelated superimposition of bursts of gravitational waves
will be well approximated, for the central limit theorem, by the
 superimposition of redshifted gaussian distributions. Therefore the power
 spectral density of the signal can be estimated by the integral
\begin{equation}
     \langle\tilde h^{2}(\omega)\rangle_{T}\approx\frac{8\,G}{c^2\,\omega^2}\int_{0}^{z_{max}}
    \frac{e^{-\frac{\left[(1+z)\omega-\omega_{0}\right]^2}{2\sigma_\omega^{2}}}}{\sqrt{2\pi}\,\sigma_\omega}
    \,\frac{R^{jet}_{GRB}(z)\,E_{GW}}{(1+z)^2\,H(z)}\,dz
    \label{eq:back4}
\end{equation}
where $\omega_0$ will the first momentum of the distribution of
the frequencies $\bar\omega(M_{BH})$ given by Eq.\ \eqref{eq:GW2},
namely $\omega_0=\int_{M_{1}}^{M_{2}}\bar
\omega(M_{BH})\,\psi(M_{BH})\,dM_{BH}$ where $\psi(M_{BH})$ is the
normalized mass distribution function for the black holes, and
$\sigma_\omega$ the second momentum. In the lack of any better
guess we will assume that $\psi(M_{BH})$ is flat for
$4\,M_\odot\le M_{BH}\le 14\,M_\odot$, thus having
$\omega_0=2\pi\,10^{3}\;\mathrm{rad}\,\mathrm{s}^{-1}$ and
$\sigma_\omega=2\pi\,200 \;\mathrm{rad}\,\mathrm{s}^{-1}$.

Finally we have reported in Fig.\ \ref{fig:fig2_GW}, the Rayleigh
power of the stochastic signal from the cosmological background
for $\eta_{GW}=0.01$.  For comparison we have also reported the
noise expected in the VIRGO experiment \cite{1998grwa.conf..524C}
averaged over one year of integration time.

\section{Discussion}\label{sect:concl}

The two central results of this paper are illustrated by Fig.\
\ref{fig:neut_back} and Fig.\ \ref{fig:fig2_GW}, where estimates
of the neutrino and gravitational waves diffuse background are
reported. the diffuse neutrino background has been calculated
accurately including many details that were not considered in
previous calculations. However the final estimate given in this
paper is in very reasonable agreement with those of previous
calculations, indicating that the order of magnitude predictions
were rather robust. The comparison of this predictions with the
best upper limits obtained so far by AMANDA for the northern sky,
indicates that an increase in sensitivity of 2-3 order of
magnitude will be required in order to set a stringent constraint
on the more plausible models of neutrino emission by cosmological
GRB's. On the other side the detection of extremely powerful
bursts of neutrinos in a km$^2$ detector cannot be ruled out,
given the extreme variability of the energetic among the various
bursts.

In \S\ref{sect:GW} it has been shown that the GW emission from
single bursts at cosmological distances, if the $\gamma$-ray
prompt emission is beamed with a small angle as suggested by
afterglow observations, is expected to be $d_L\,\tilde h\lesssim
10^{-27}\;\mathrm{Gpc}/\sqrt{\mathrm{Hz}}$ which is well below the
detection threshold of presently planned experiments. However if
the emission is beamed only a small fraction (one over 300) of the
GRB's are observed by $\gamma$-ray satellites. This implies that
small amplitude pulse trains of GW's impinge over the detector at
a frequency that could be as high as half per minute, which is
practically a continuous signal. Even if the individual pulse is
small, integrating over a reasonable observation time (order of
one year) an excess Rayleigh power should emerge from the
instrumental noise. The frequency spectrum of this eventual excess
should give a direct information on the characteristic time scale
of the collapse and on the cosmological evolution of the GRB rate
in the recent past ($z\approx 1-5$). The predicted amplitude is
conservatively estimated to be in the range of $5\times
10^{-26}\;1/\sqrt{\mathrm{Hz}}$ averaging the Rayleigh power over
one year. This estimate is rather robust because does not depends
on the beaming factor and depends only slightly from the large
scale distribution of the GRB sources. The cosmological origin of
the excess noise could be proved by detecting a dipole anisotropy.
In addition the auto correlation spectrum of the noise shall carry
an imprint of the characteristic duration of the GW pulse trains.
The detection of those two signatures, even if perhaps not
possible with presently planned experiments, can give the
important additional evidence of the cosmological origin of the
stochastic signal and informations on the physics of GRB's.

\end{document}